\documentclass[12pt]{article}
\usepackage{color}

\usepackage{jcappub_mod}
\usepackage{amsthm}

\usepackage{multirow} 
\usepackage{bm}
\usepackage{amsmath}
\usepackage{color}
\usepackage{natbib}
\usepackage{graphicx} 

\makeatletter
\renewcommand\footnoterule{%
  \vspace{-5pt}
  \kern-3\p@\hrule\@width.4\columnwidth%
  \kern10\p@}
\makeatother

\def\be{\begin{equation}}
\def\ee{\end{equation}}
\def\ba{\begin{eqnarray}}
\def\ea{\end{eqnarray}}

\newcommand{\Mpl}{M_{\rm Pl}}

\newcommand{\lmax}{\ell_{\mathrm{max}}}

\definecolor{darkgreen}{cmyk}{0.85,0.2,1.00,0.2} 
\definecolor{purple}{cmyk}{0.5,1.0,0,0}

\newcommand{\ModeCode}{{\sc Mode\-Code}}
\newcommand{\MultiNest}{{\sc MultiNest}}

\newcommand{\CosmoMC}{{\sc CosmoMC}}

\begin{document}
\title{Planck Constraints on \\ Monodromy Inflation}

\author[1]{Richard Easther}
\author[2,3]{and Raphael Flauger}

\affiliation[1]{Department of Physics, University of Auckland, Private Bag 92019, \\ Auckland, New Zealand}
\affiliation[2]{School of Natural Sciences, Institute for Advanced Study,\\ Princeton, NJ 08540, USA}
\affiliation[3]{Center for Cosmology and Particle Physics, Department of Physics,\\
New York University, New York, NY, 10003, USA}
\emailAdd{r.easther@auckland.ac.nz}
\emailAdd{flauger@ias.edu}

\date{\today}

\abstract{We use data from the nominal Planck mission to constrain  modulations in the primordial power spectrum associated with monodromy inflation. The largest improvement in fit relative to the unmodulated model has $\Delta\chi^2\approx 10$ and we find no evidence for a primordial signal, in contrast to a previous analysis of the WMAP9 dataset, for which $\Delta\chi^2\approx 20$.  The Planck and WMAP9 results are broadly consistent on angular scales where they are expected to agree as far as best-fit values are concerned.  However, even on these scales the significance of the signal is reduced in Planck relative to  WMAP, and is consistent with a fit to the ``noise'' associated with cosmic variance. Our results motivate both a detailed comparison between the two experiments and a more careful study of the theoretical predictions of monodromy inflation.}

\maketitle

\section{Introduction} \label{sec:intro}

After the first cosmological data release from the Planck satellite, all observations remain  consistent with a  $\Lambda$CDM universe in which the perturbations were sourced by the vacuum fluctuations of the inflaton \cite{Ade:2013zuv,Ade:2013uln}. The small observed departure from scale-invariance is in agreement with the generic predictions of simple single-field inflationary scenarios, as is the apparent adiabaticity  and Gaussianity  \cite{Ade:2013ydc} of the perturbations.   

Simple, single-field inflationary models  predict that a  stochastic gravitational wave (tensor) background is generated by  quantum fluctuations of  spacetime in the primordial universe. It was shown by Lyth~\cite{Lyth:1996im} that the amplitude of any background of tensor perturbations is correlated with the total excursion of the inflaton field.  Observationally, there is currently no evidence for the existence of such a background \cite{Ade:2013zuv}, but any detectable signal would imply that the inflaton field varies over a super-Planckian range. 

If the relevant physics is well-described by a generic effective field theory with a sub-Planckian cut-off, the potential is unlikely to be smooth over super-Planckian field ranges \cite{Lyth:1996im}.  Consequently, in scenarios with super-Planckian inflaton expectation values, one expects that symmetries ensure  the flatness of the potential and that the vacuum expectation value of the inflaton does not affect the masses of any fields to which it is coupled. The role of symmetries to protect the flatness of the potential was discussed early on in the context of chaotic inflation~\cite{Linde:1983gd,Linde:1987yb}, and pseudo-Nambu-Goldstone bosons were introduced as an inflaton candidate in~\cite{Freese:1990rb} to ensure naturalness. 

The best motivated scenarios with super-Planckian inflaton expectation values are those in which  shift-symmetries naturally arise within candidate theories of fundamental physics, such as string theory.  A promising string theoretic model with super-Planckian field excursions was constructed in Type IIA theory by Silverstein and Westphal~\cite{Silverstein:2008sg}. A closely related model in Type IIB string theory with better control over moduli stabilization was  proposed in~Ref.~\cite{McAllister:2008hb} by McAllister, Silverstein and Westphal. This model was studied in more detail in Ref.~\cite{Flauger:2009ab}. In both cases the potential  ``flattens out'' relative to a quadratic potential at large field values, which is a generic feature of back-reaction~\cite{Dong:2010in}. Intriguingly, inflationary potentials with the asymptotic form $V\sim \phi^p$ with $p<2$  are a good fit to recent Planck data \cite{Ade:2013uln}.

The large field range in these scenarios is achieved with the help of a periodic direction in field space which exhibits a shift symmetry to all orders in string perturbation theory.  This periodic direction is ``unwrapped'' by placing branes on appropriate cycles in the internal space, so that  the field trajectory may be visualized as a spiral.  
Distinct  points on the spiral that project to the same point on the circle are distinguished by a charge or energy that grows with each completed circuit. This change of the configuration as the field moves round the circle is known as monodromy, and the associated cosmological model is {\em monodromy inflation\/} \cite{Silverstein:2008sg,McAllister:2008hb,Flauger:2009ab}.    

Monodromy models thus break time translation invariance on two distinct scales, one associated with the frequency of the periodic motion, and one associated with the velocity of the inflaton field $\phi$ averaged over one period.  These scales may both be reflected in the power spectrum, combining the weak scale-dependence seen in almost all inflationary models models with a rapid modulation \cite{McAllister:2008hb,Flauger:2009ab}, driven by the inflaton's movement around the  circular direction. The amplitude of this modulation is model-dependent and can have an arbitrarily small amplitude so that there is no guarantee of a signal in the data. However, the detection of an appropriately modulated power spectrum in combination with the scalar spectral index and tensor amplitude predicted by monodromy inflation would provide dramatic support for the model itself and the  theoretical framework from which it was derived.  Depending on the couplings of the inflaton to other degrees of freedom, the model can give rise to further interesting phenomenology~\cite{Amin:2011hj,Barnaby:2011qe}.

Recently, Ref.~\cite{Peiris:2013opa} examined the WMAP9 dataset  \cite{Bennett:2012fp,Hinshaw:2012fq} and found tentative evidence for an oscillatory power spectrum, in which the modulation had a relatively large amplitude and high frequency.  In this paper we search for an oscillatory power spectrum in the Planck dataset. Unlike Ref.~\cite{Peiris:2013opa} which employed the combination of \ModeCode\ \cite{Mortonson:2010er,Easther:2011yq}, \MultiNest\ \cite{Feroz:2007kg, Feroz:2008xx} and \CosmoMC\ \cite{Lewis:2002ah} for parameter estimation and the computation of Bayesian evidence,  the calculation in this paper is implemented via a grid search as in~\cite{Flauger:2009ab}, using the analytic approximation to the monodromy inflation power spectrum derived in~\cite{Flauger:2009ab}, and caching the transfer functions needed to compute the angular power spectrum.  This approach is much faster because it skips the numerical calculation of the power spectrum, and it permits a brute-force search for a modulated signal in the microwave background as well as computation of Bayesian evidence. However, it relies on the accurate analytic expression for the power spectrum from~\cite{Flauger:2009ab} and makes use of the fact that the parameters of the primordial power spectrum are not strongly degenerate with the other LCDM parameters as well as foreground, calibration, and beam parameters. It is thus not applicable to scenarios where an analytic form of the power spectrum is not known or where degeneracies with these parameters exist. 

This paper is structured as follows. In Section~\ref{sec:monod} we outline the relevant features of the monodromy model. In Section~\ref{sec:constr} we present the parameter estimates and discuss the best fit models for the Planck dataset. We estimate the change in the log-likelihood expected from fitting a rapidly modulated power spectrum to Gaussian noise, showing that it is not well-described by the usual $\chi^2$ distribution. We briefly turn to model selection and compute the Bayesian evidence before discussing our results in Section~\ref{sec:disc}. 

\section{Mondromy Inflation: Potential and Power Spectrum \label{sec:monod}} 

Our analysis relies on the low energy effective field theory for axion monodromy inflation which, besides gravity, contains a single scalar degree of freedom. We will discuss the limitations of this effective field theory in Section~\ref{sec:disc}. The scalar is minimally coupled to gravity and has a canonical kinetic term. The potential is
\be
V(\phi)=V_0(\phi)+\Lambda^4\cos\left(\frac{\phi}{f} +\psi\right)\,,
\ee
where the decay constant $f$ is an energy scale that reflects the periodicity of the underlying compact scalar field \cite{McAllister:2008hb,Flauger:2009ab}. Additional higher harmonics are possible, but these are small in the region of parameter space in which we have good theoretical control. The scale $\Lambda$ decreases exponentially with the volume of certain cycles in the internal space. Oscillatory features in the power spectrum thus arise naturally in these models, but can easily be undetectably small. Finally, $\psi$ is a constant phase. In general, $V_0(\phi)$ is a slow-roll potential that is well approximated by 
\be
V_0(\phi)=\mu^{4-p}\phi^p\qquad\text{with}\qquad p<2\,,
\ee
during inflation and is quadratic near the minimum. The energy scale $\mu$ is not predicted by the underlying theory, but is determined from the overall amplitude of the primordial perturbations. Such large field models predict an amplitude of the primordial gravitational wave spectrum near current observational  limits.  In this analysis we restrict attention to the linear potential ($p=1$), and parameterize the potential as
\be
V(\phi) = \mu^3 \left[\phi + bf  \cos\left(\frac{\phi}{f} + \psi \right)  \right]  \label{eq:v}  \, .
\ee 

As the universe expands, the physical momentum of each inflaton mode redshifts, and the modes undergo parametric resonance when the background frequency is roughly twice their natural frequency, $\dot\phi/f\approx 2k/a$. The resulting excitations give rise to the oscillatory power spectrum. The spectrum for the linear potential was derived in~\cite{Flauger:2009ab} 
\be \label{eq:delta} 
\Delta_\mathcal{R}^2(k)=\Delta_\mathcal{R}^2(k_\star)\left(\frac{k}{k_\star}\right)^{n_s-1}\left[1+\delta n_s\cos\left(\frac{\phi_k}{f}+\varphi\right)\right]\,,
\ee
where $\phi_k$ denotes the value of the scalar field when the mode with comoving momentum $k$ exits the horizon, $\varphi$ is some phase that encodes both inflationary physics and the  unknown mapping to present-day scales, which depends on the detailed expansion history of the post-inflationary universe \cite{Liddle:2003as,Peiris:2008be,Adshead:2010mc,Mortonson:2010er,Easther:2011yq}, and
\be  \label{eq:ns} 
\delta n_s=\frac{12b}{\sqrt{(1+(3f \phi_\star)^2)}}\sqrt{\frac{\pi}{8}\coth\left(\frac{\pi}{2 f\phi_\star}\right) f\phi_\star}\,,
\ee
where $\phi_\star$ denotes the value of the scalar field when the pivot scale exits. The calculation was extended to general slow-roll potentials in~\cite{Flauger:2010ja} in the limit $f\phi_\star\ll1$.\footnote{See also~\cite{Behbahani:2011it} for a derivation in the context of the effective field theory of inflation~\cite{Cheung:2007st}.} We focus on the linear potential but, up to the change in the spectral index, the results can be mapped to more general models via the identification
\be
f\phi_\star=\frac{H}{\omega}\,,
\ee
where $\omega$ denotes the frequency with which the background oscillates.

The equations governing the evolution of  the field are well known, and the number of $e$-folds of inflation $N$ is connected to field values of $\phi$ by 
\be \label{eq:nval}
N= \int_{\phi_{end}}^\phi \frac{d\phi}{\sqrt{2\epsilon}\Mpl} \qquad \text{so that}\qquad  \phi_k \approx \sqrt{2N_k}\Mpl=\sqrt{2(N_\star-\ln(k/k_\star))}\Mpl  \,.
\ee
where the final approximate equality is reliable  for the values of $b$ that are of interest in our analysis. Here $\Mpl$ denotes the reduced Planck mass, which is set equal to unity in what follows. Notice that we use the square-root expression for $\phi_k$ in our analysis and do not expand it to recover the more commonly studied oscillations in log-space. This was already used for the WMAP5 analysis in Ref.~\cite{Flauger:2009ab}, but is not important for WMAP. However, for the range of multipoles probed by Planck this difference becomes noticeable.

\begin{figure}[t]
\begin{center}
\includegraphics[width=5.0in]{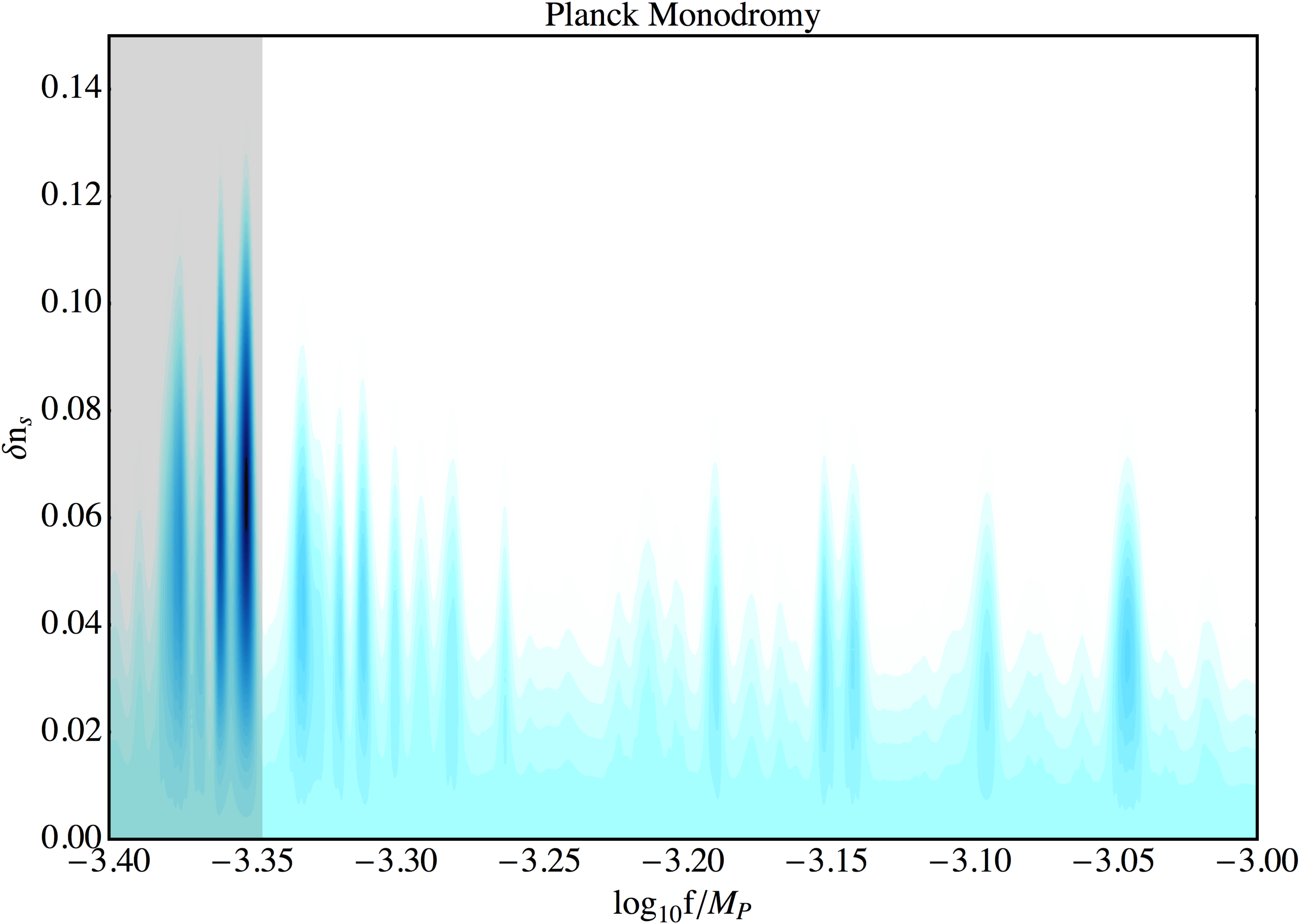}\hskip-.5cm
\includegraphics[width=4.9in]{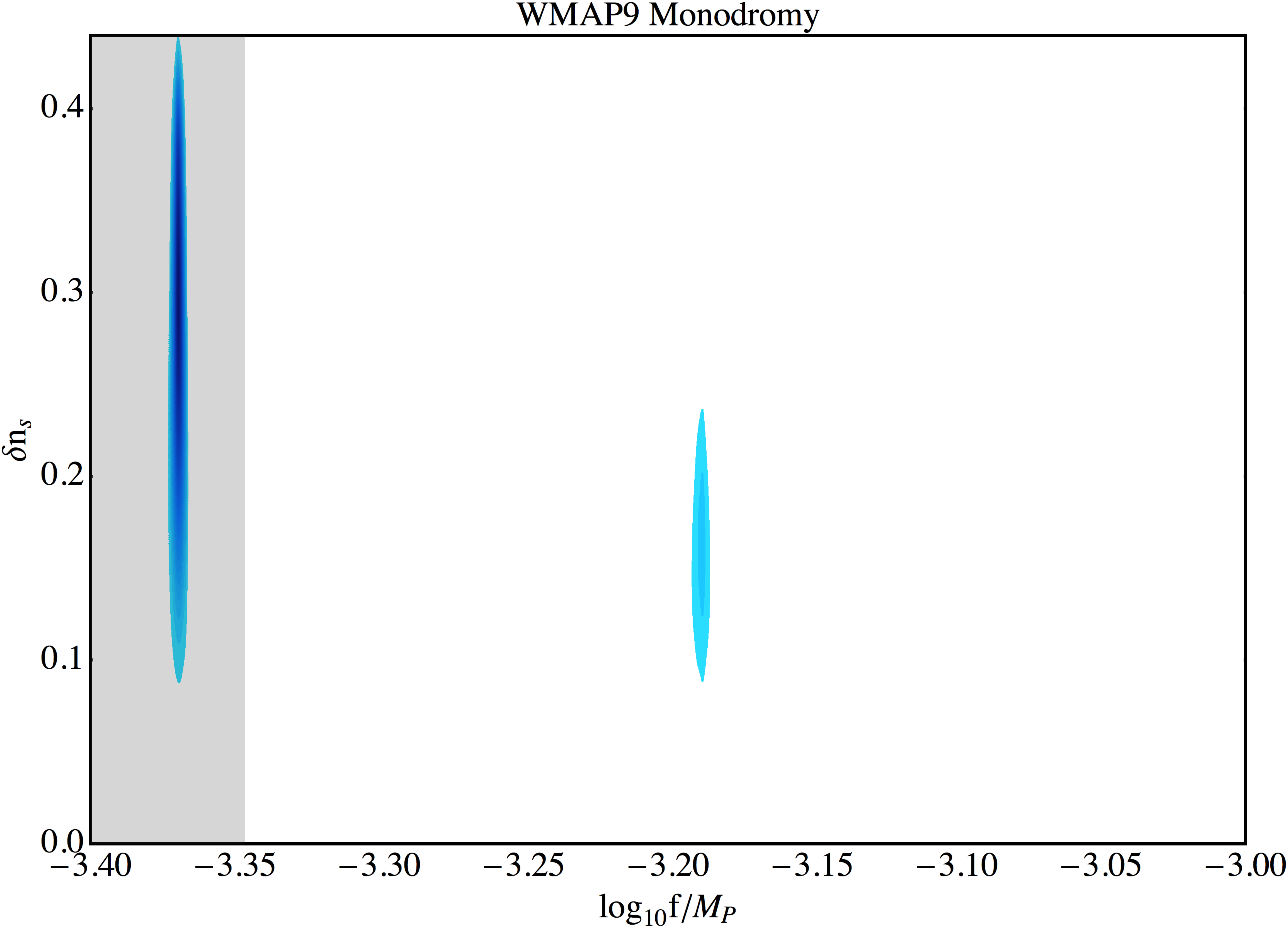}\\[.5cm]
\end{center}
\caption{\label{fig:dnsvsf34} Marginalized posterior distributions for inflationary parameters derived from Planck (top) and WMAP (bottom). The WMAP plot reproduces the results of~\cite{Peiris:2013opa} using the methods and notation of the present analysis. The Planck analysis is performed over the full $\ell$ range, combined with the WMAP polarization data on large scales. The shaded region indicates values of $f$ for which the single field effective field becomes strongly coupled and a more careful study of the underlying stringy model may be necessary to see if the predictions of the model change qualitatively.}
\end{figure}

\section{Constraints from Planck}\label{sec:constr}

Monodromy models have previously been constrained  using the  WMAP5 \cite{Flauger:2009ab}, WMAP7  \cite{Meerburg:2011gd,Aich:2011qv} and, most recently,  the final WMAP9 \cite{Peiris:2013opa} datasets. In Ref.~\cite{Peiris:2013opa},  the parameter space was explored with \MultiNest, the Mukhanov-Sasaki equation describing the evolution of the inflaton modes was solved directly with \ModeCode, and the angular power spectrum computed for each combination of parameters. The analysis found an improvement of $\Delta\chi^2_\text{eff}\sim 20$ in the match to the power spectrum at the best-fit values of $f$ and $b$, and it is natural to ask whether the same improvement is seen in the Planck data.   

To answer this question we adopt a different strategy from the one in reference~\cite{Peiris:2013opa}, working directly with equations~(\ref{eq:delta}) and~(\ref{eq:ns}), and computing the CAMspec, Commander, and lowlike likelihood functions for each point on a grid in the parameter space spanned by $\mu$, $\delta n_s$, $f$, and the phase $\varphi$ as in~\cite{Flauger:2009ab}. For most runs, the other ``background'' cosmological parameters are set to their best-fit values derived from the Planck data for the unmodulated model. Because only the parameters in the power spectrum change, the transfer functions are only evaluated once, accelerating the computations significantly. If only one or two of the background parameters are varied, the method generalizes trivially by assigning different values of the background cosmological parameters to different cores so that the transfer function is still only calculated once on each of the cores. We use this to vary $\Omega_b h^2$ in additional runs for the region $f>10^{-2}\Mpl$ and to check that no significant degeneracies exist between the power spectrum parameters and $\Omega_bh^2$ near the best-fit points in the region of smaller axion decay constants. For a complete analysis varying all cosmological parameters as well as foreground, calibration, and beam parameters, a grid search is certainly intractable and we make use of the fact that there are no strong degeneracies between these parameters and the parameters of interest.

\begin{figure}[t]
\begin{center}

\includegraphics[width=5.1in]{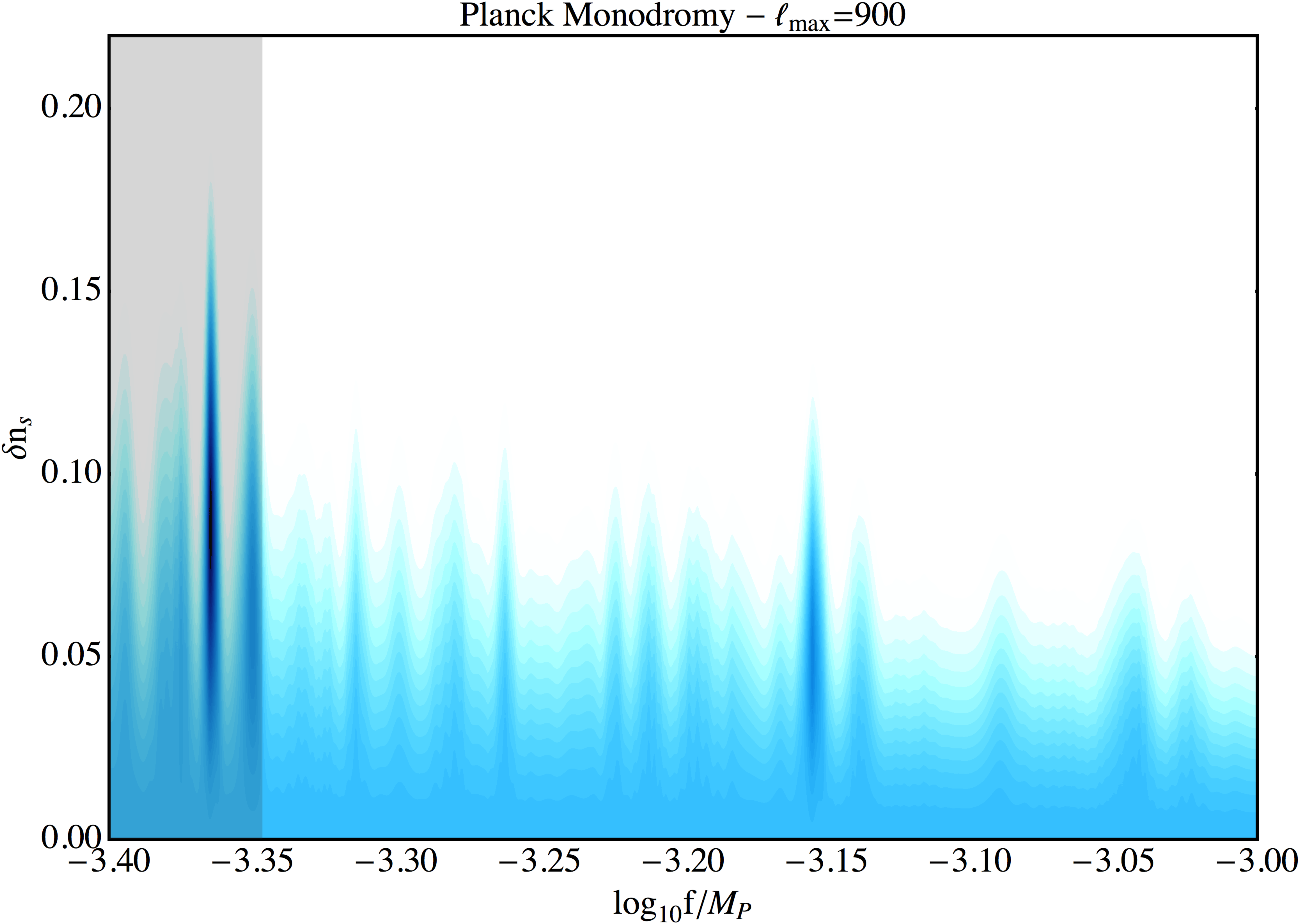} 
\\[1.5cm]
\includegraphics[width=5.1in]{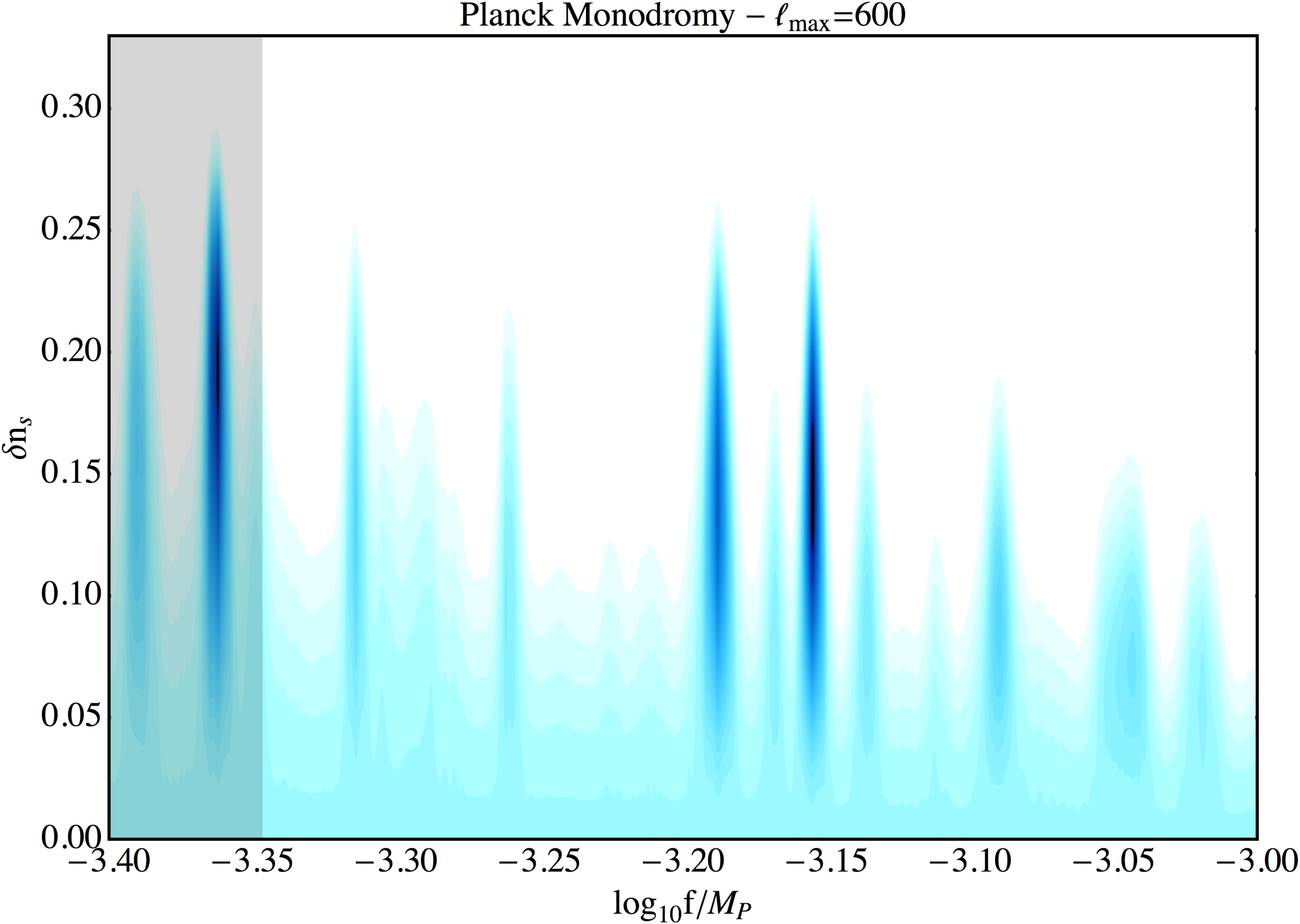}
\end{center}
 \caption{\label{fig:dnsvsf34LMAX} Marginalized posterior distributions for inflationary parameters from Planck with $\lmax$ of 900 (top) and 600 (bottom).  }
\end{figure}

Our grid consists of $16$ points for the amplitude of scalar perturbations $\Delta_\mathcal{R}^2$, $16$ points for the amplitude of the modulation, $32$ points for the phase, with $400$ logarithmically spaced points for $4\times10^{-4}\Mpl<f<10^{-3}\Mpl$ and $1000$ logarithmically spaced points for $10^{-3}\Mpl<f<10^{-1}\Mpl$. We do not consider axion decay constants below $f=4\times10^{-4}\Mpl$ because the effective field theory with a single scalar degree of freedom is only weakly coupled provided $\omega\ll 4\pi f$~\cite{Behbahani:2011it}. For axion decay constants which lead to oscillations in the background geometry with frequencies above $\omega\approx 4\pi f$, which corresponds to decay constants below $f\approx4.5\times10^{-4}\Mpl$, other degrees of freedom in the underlying stringy model become important and conclusions derived from the single field effective field theory gradually become unreliable. This region is shaded in the plots. We also impose an upper bound on the axion decay constant of around $f<10^{-1}\Mpl$. For larger values it is very likely to be impossible to embed the effective field theory into string theory~\cite{Banks:2003sx,ArkaniHamed:2006dz}. In the runs in which we varied the baryon content, we used $16$ points for $\Omega_b h^2$.

Figure~\ref{fig:dnsvsf34}  shows the constraints on $f$ and $\delta n_s$ after marginalizing over the remaining paramters, $\Delta_\mathcal{R}^2$ and $\varphi$, derived from the Planck+WP\footnote{The Planck+WP data combines the lowlike likelihood derived from Planck temperature data for the nominal mission with the WMAP polarization data on large scales~\cite{Planck:2013kta}, with the Commander and CAMspec likelihoods.} and WMAP9 dataset, for the range of frequencies  within the regime of validity of the single field effective field theory that was not studied in the Planck paper~\cite{Ade:2013uln}. The best-fit frequency found in this range of axion decay constants is close to the one found in the WMAP9 dataset. However, the modulation amplitude is significantly smaller and one finds a substantially smaller improvement of $\Delta\chi^2_\text{eff}\approx 7.8$. Consequently, the immediate conclusion is that the strong signal seen in Ref.~\cite{Peiris:2013opa} is not recovered from the Planck dataset. 

The best-fit value of the axion decay constant in this range is close to the value for which the single field effective field theory becomes unreliable. Consequently, a more careful theoretical study of the predictions of axion monodromy inflation with a small axion decay constant is needed to decide whether we have constrained axion monodromy inflation, or simply provided a phenomenological bound on oscillations in the power spectrum of the form~\eqref{eq:delta}, independently of its physical motivation. 

A natural explanation for the observed decrease in significance between WMAP and Planck would be that both data sets lead to similar improvements where they overlap and are cosmic variance limited, but that the modulation no longer provides a good fit to the data on the small angular scales measured only by Planck. 
It is thus natural to check whether the large improvement in $\Delta \chi^2$ is recovered when only the low-$\ell$ portion of the Planck power spectrum is used.  
\begin{figure}[tbp]
\begin{center}

\mbox{}

\includegraphics[width=5in]{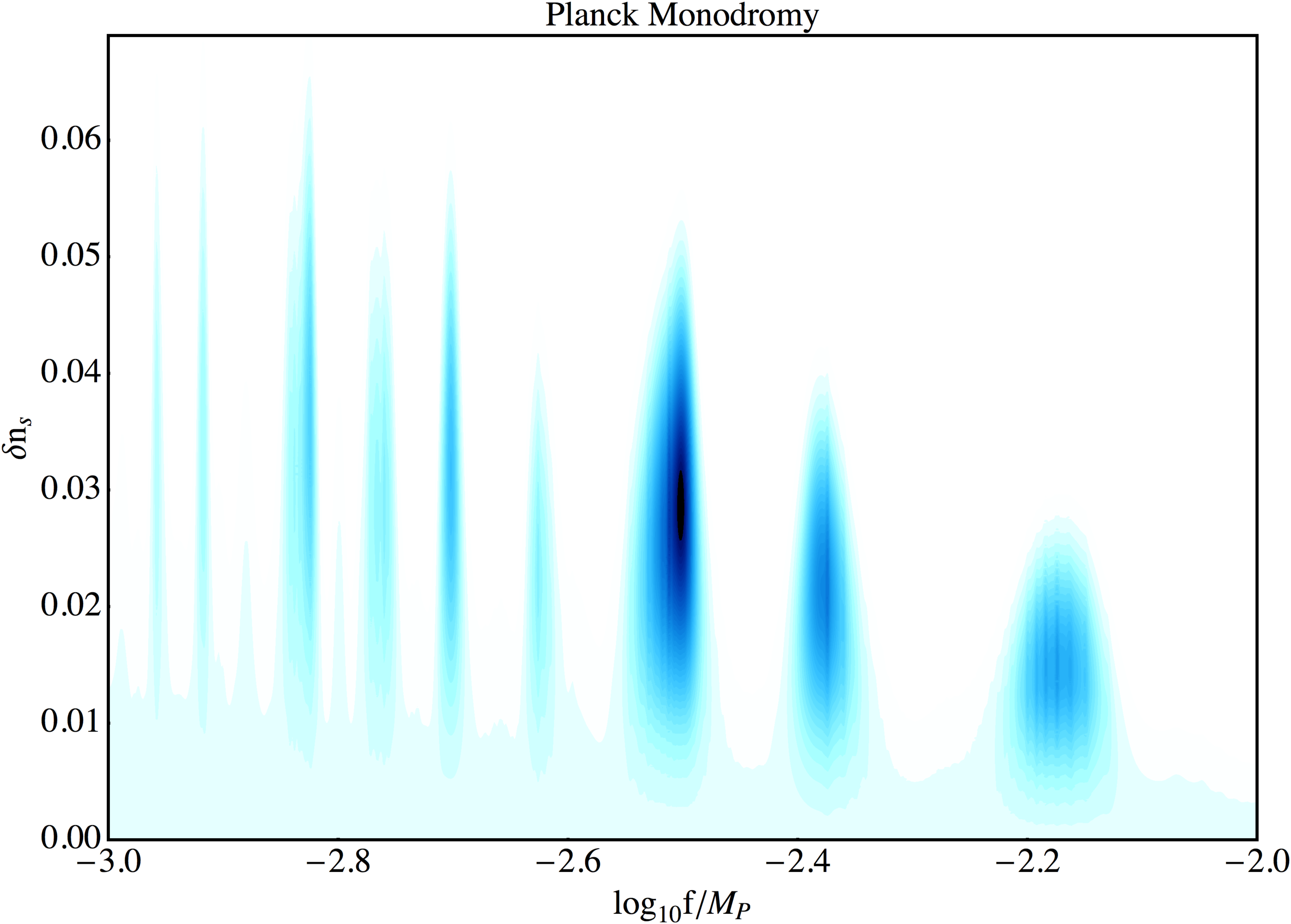} 
\\[.7cm]
\includegraphics[width=5in]{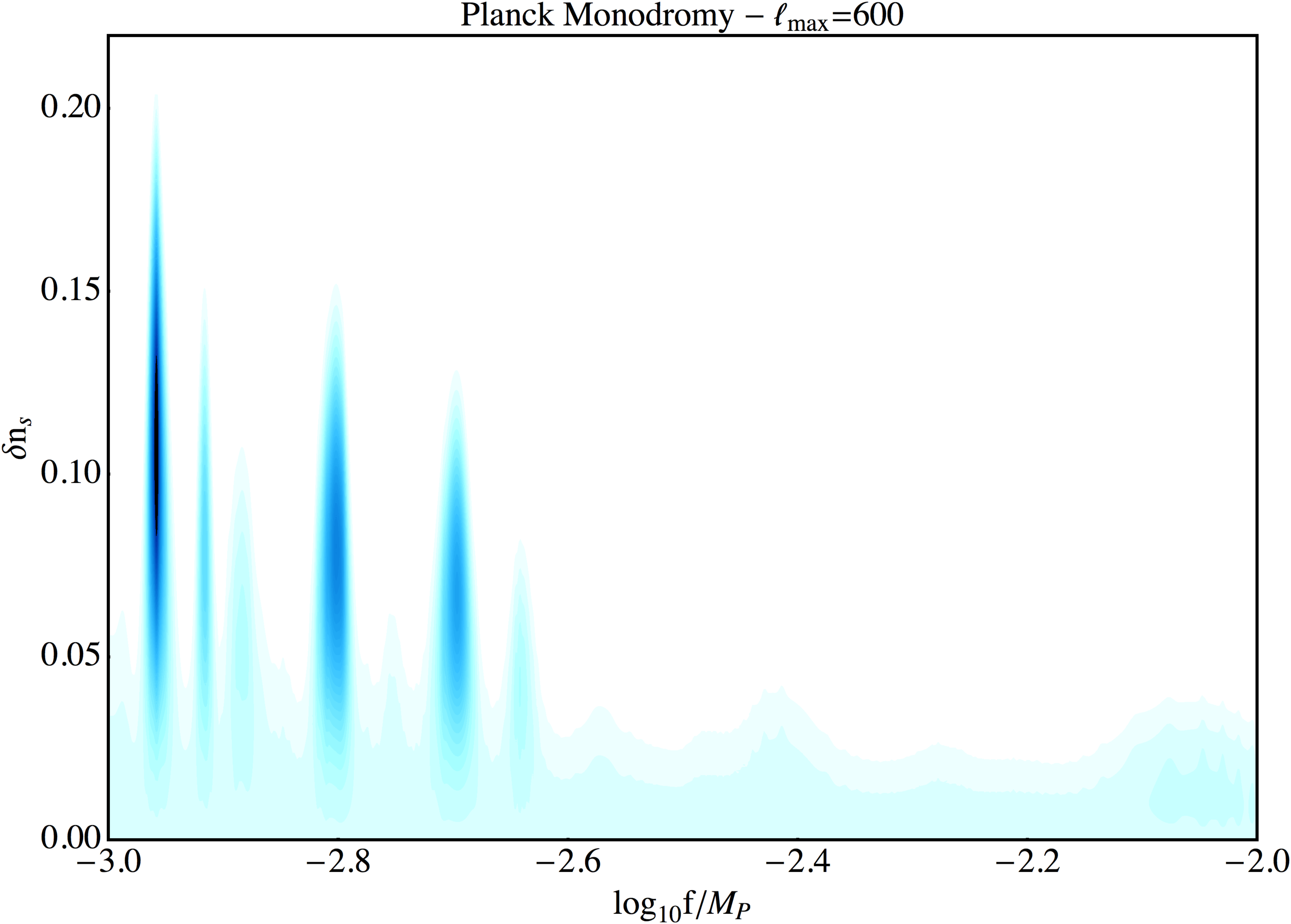}\hskip-.5cm
\end{center}
\caption{\label{fig:dnsvsf3LMAX} Marginalized posterior distributions for inflationary parameters for larger values of the axion decay constant. The full angular power spectrum is used in the top panel, while the lower panel has $\lmax= 600$.   }
\end{figure}
In Figure~\ref{fig:dnsvsf34LMAX} we show the Planck constraints for $\lmax$ of 600 and 900. In both cases we see a prominent feature at $\log_{10}(f) \approx -3.37$, as in WMAP9\footnote{The small shift in the best-fit value for the axion decay constant relative to Ref.~\cite{Peiris:2013opa} is due to the small changes in the parameters of the background cosmology between Planck and WMAP.} and the best-fit modulation amplitude in Planck approaches that of WMAP9 as $\ell_\text{max}$ is lowered. However, the best-fit for Planck still has $\Delta \chi^2\approx 10$, which is reduced relative to WMAP, which had  $\Delta \chi^2\approx20$.  Consequently, even at these scales, there are significant differences between the Planck and WMAP likelihoods. We have not yet been able to pinpoint the origin of this discrepancy, but discuss it in more detail in Section~\ref{sec:disc}. 

\begin{figure}[t]
\begin{center}
\includegraphics[width=6.2in]{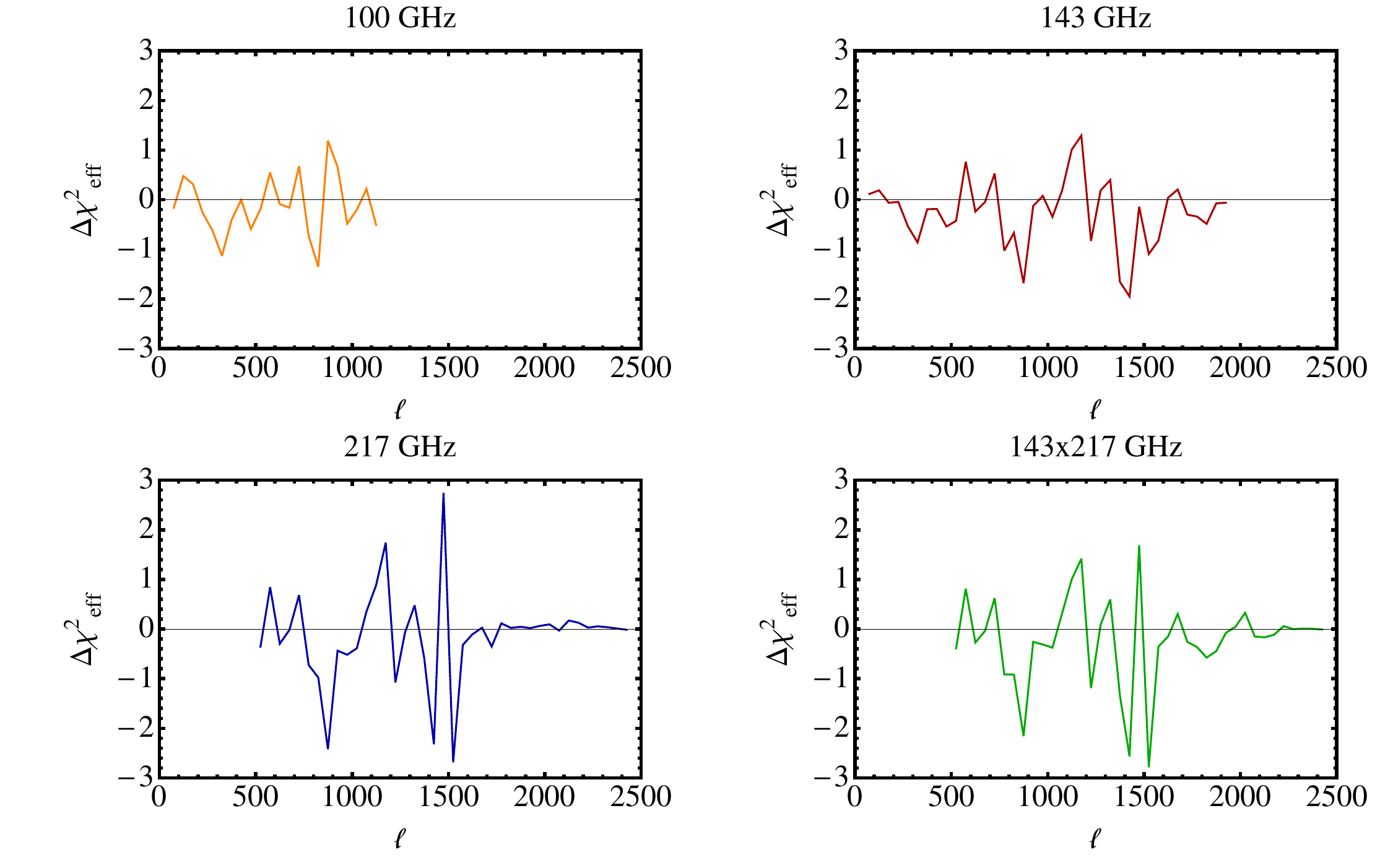}\\[.4cm]
\includegraphics[width=6.2in]{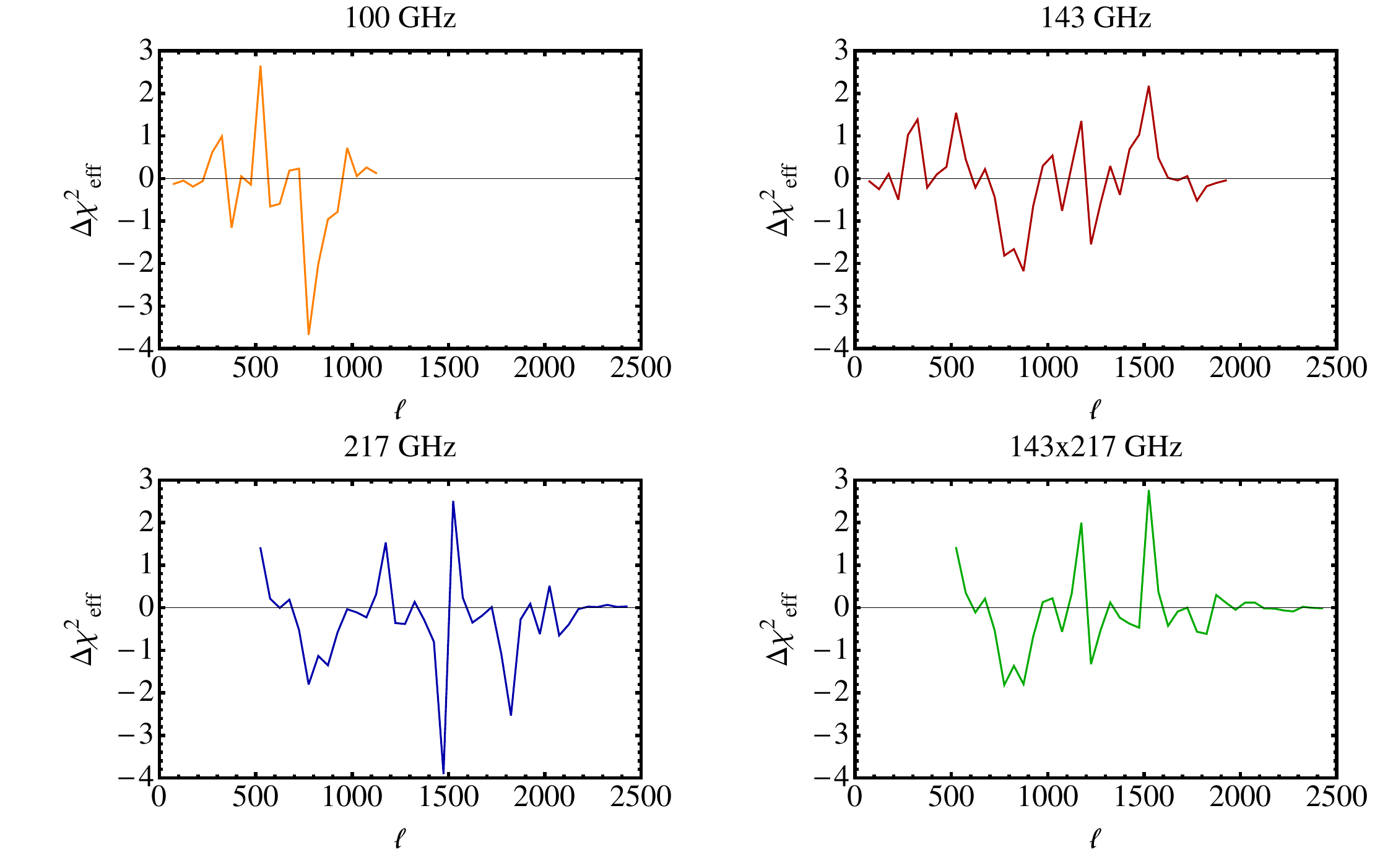}
\end{center}
\caption{\label{fig:dchi2best}Difference in $\chi^2_\text{eff}$ between the best-fit point near $f=4.4\times10^{-4}\Mpl$ and $f=3.1\times 10^{-3}\Mpl$ in the top and bottom half of the Figure derived from the blocks of the CAMspec covariance matrix for the different frequency channels. }
\end{figure}

As a useful additional diagnostic, we show the difference $\Delta\chi^2_\text{eff}=-2\Delta\ln\mathcal{L}$ broken up into bins of $\Delta\ell=50$ between the feature model with $\log_{10}(f) \approx -3.37$ and the smooth reference model for the frequency channels used in CAMspec in the top half of Figure~\ref{fig:dchi2best}. The plots use the covariance matrices for the individual frequency channels. We see that there is good agreement between the different frequency channels as one would expect for a primordial signal. An exception is a small range of multipoles around $\ell\approx 900$ where the fit is worse for the modulated spectrum in the 100 GHz channel but better for the others. Even though the noise becomes important for lower multipoles in $100$ GHz than in $143$ and $217$ GHz due to the larger beam, it should still be subdominant on these angular scales. So this small glitch may deserve a closer look. One might speculate that this is related to the fact that the mask used for the $100$ GHz maps is different from that used for the $143$ and $217$ GHz maps, but we have not studied this in detail. Overall, it is the agreement between the different channels that stands out rather than the small disagreement.

Another natural question is whether there is evidence for modulations with lower frequencies (larger $f$), which are more effectively constrained in a dataset with a large dynamic range such as Planck. Figure~\ref{fig:dnsvsf3LMAX}  shows the results of a search for modulations at larger values of $f$, $10^{-3} \Mpl \le f \le 10^{-2} \Mpl$, which overlaps with the analysis of Ref.~\cite{Ade:2013uln}. For the default multipole ranges of CAMspec with $\ell_\text{max}=2500$ we see a number of localized peaks in the posterior consistent with those found in Ref.~\cite{Ade:2013uln}. The best-fit in this range is also the best fit over the full range of $4\times10^{-4} \Mpl \le f \le 10^{-1} \Mpl$ with $\Delta\chi^2_\text{eff}=9.8$. Limiting the highest multipole to $\lmax =600$, this peak is entirely absent, indicating that this modulation fits a feature in the high-$\ell$ portion of the angular power spectrum. This is also seen in the bottom half of Figure~\ref{fig:dchi2best}. Much of the improvement in the fit comes from the fifth and sixth acoustic peak in the $217$ GHz data without similar improvements in the $143$ GHz or $143\times 217$ GHz data. This dependence on the frequency and the fact that these features around the fifth and sixth acoustic peak in the 217 GHz map are seen in detector set correlations but not in survey cross-correlations~\cite{cleaning}, suggests that this is most likely not a primordial signal. Finally, there are no significant improvements in the remaining range of axion decay constants $10^{-2} \Mpl \le f \le 10^{-1} \Mpl$ with $\Delta\chi^2_\text{eff}\approx3$.

\begin{figure}[t]
\begin{center}
\includegraphics[width=5.5in]{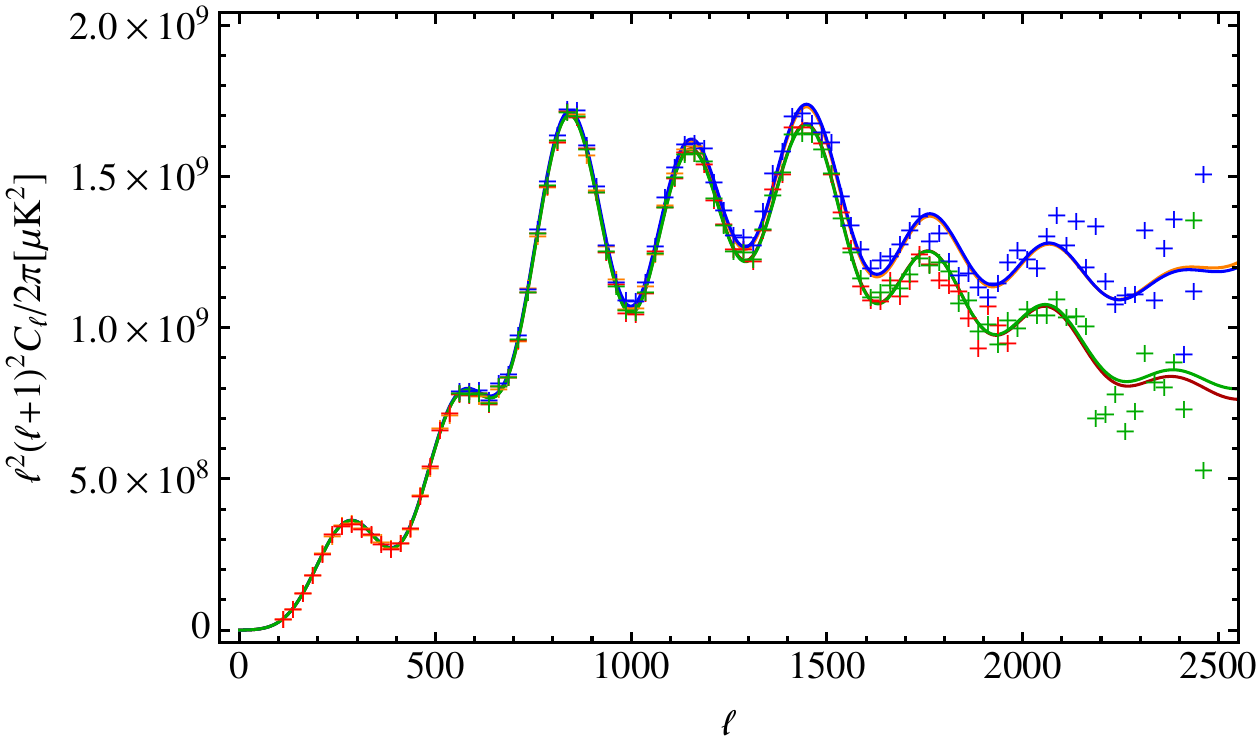}
\end{center}
\caption{\label{fig:camspec}Comparison of data used in CAMspec binned with $\Delta\ell=25$ with the smooth best-fit model (including the model for extragalactic foregrounds). The $100$ GHz data is shown in orange, the $143$ GHz data in red, the $217$ GHz data in blue, and finally the $143\times 217$ cross spectra in green.}
\end{figure}

\begin{figure}[t]
\begin{center}
\includegraphics[width=5.0in]{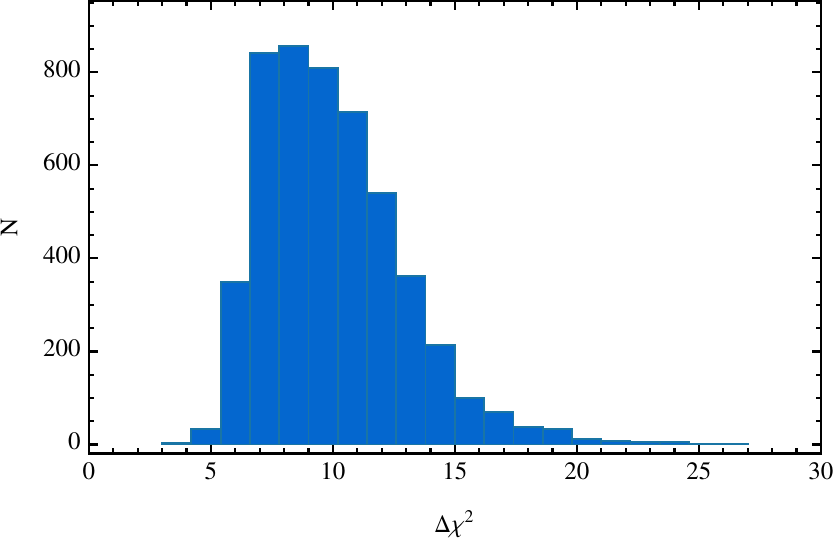}  
\end{center}
\caption{\label{fig:histo}    The best fit $\Delta \chi^2$ obtained from fitting a modulated model to Gaussian noise with 5000 trials. A typical ``fit to noise" has a $\Delta \chi^2 \sim 10$. }
\end{figure}

Let us briefly discuss the significance of the values of $\Delta\chi^2_\text{eff}$ found in this and previous analyses. One naively expects $\Delta \chi_\text{eff}^2\sim 1$ for each parameter that is added to the model, simply by providing a better fit to the noise. Consequently, an improvement of $\Delta \chi^2_\text{eff}\gtrsim 10$ appears potentially significant given that the modulation is described by three parameters, phase, amplitude and frequency. However,  this rule only applies to parameters that enter linearly into the fit, or parameters we can linearize in near the best-fit point.  However,  the axion decay constant varies over too large a range for these conditions to hold and we must account for the look-elsewhere effect.  Using  simulations we derive the distribution of best-fit $\Delta\chi^2$ in the absence of a signal.\footnote{This topic can be treated analytically, and will be discussed in detail in a future publication.}
 We will assume that the best-fit smooth model has been subtracted from the data, or equivalently that we use it as a reference to compute  $\Delta\chi^2$. We make the simplifying assumption that the residual is uncorrelated random Gaussian noise with an amplitude given by the expected cosmic variance and noise. This ignores effects of the mask as well as correlations in the noise, but provides a good first approximation. We fit the modulation to this residual noise, varying the same parameters as in the real analysis. The distribution of the largest improvement in each of the simulations  is shown for 5000 simulations in Figure~\ref{fig:histo}. We see that improvements of $\Delta \chi^2\sim 10$ are typical. Consequently, the modulated monodromy spectrum yields an improvement consistent with a fit to noise for all the searches performed with the Planck dataset reported here.

Since we have computed the likelihood function for every point on our grid in parameter space, we can pose this as a model selection problem by computing Bayesian evidence for the model and comparing it to the unmodulated LCDM case.  With  uniform priors\footnote{These results are derived for a prior of $0<b<0.5$, $-\pi<\phi<\pi$, $4\times 10^{-4}\Mpl<f<10^{-1}\Mpl$ for the modulated model and $0.95<n_s<0.98$ for LCDM, effectively independent of the prior for the amplitude of scalar perturbations as it appears in both models.  } the evidence is
\begin{equation}\label{eq:flatprior}
E=\frac{1}{\text{Vol}_M}\int d {\boldsymbol\alpha}\, \mathcal{L}(\alpha_i)\,,
\end{equation}
where  $\text{Vol}_M$ denotes the volume in parameter space allowed by the prior, $\alpha_i$ are the various model parameters, and $\mathcal{L}$ is the likelihood function. Keeping the background cosmology fixed and integrating over the parameters of the power spectrum of scalar perturbations, we find $\Delta\ln E\approx 1.5$ or betting odds in favor of LCDM of $20:1$.

Lastly, monodromy inflation is known to give rise to resonant non-Gaussianity \cite{Chen:2008wn,Flauger:2010ja}, and the form of the non-Gaussianity is correlated with the modulation seen in the power spectrum. Naively applying the effective field theory prediction to the best-fit point with $f=4.4\times 10^{-4}\Mpl$ one expects an amplitude of $f_{NL}\approx 400$. While the signal-to-noise ratio for the 3-point function is at best comparable to the one in the power spectrum~\cite{Behbahani:2011it}, hints in the 3-pt function might provide additional support for the model. However, further theoretical work is needed to understand the prediction of the model in the small-$f$ region, and the difference between WMAP and Planck should be understood. Additional support for a search of these oscillatory signals comes from models that give rise to a detectable 3-point function without a signal in the power spectrum~\cite{Behbahani:2012be}. Such a search is very difficult, but could potentially lead to extremely interesting results.
 
\section{Discussion}\label{sec:disc}

We have searched for evidence of the modulated power spectrum associated with monodromy inflation in the first Planck data release and have found no evidence for a modulated power spectrum. A previous analysis of the WMAP9 dataset found a specific modulation with $\Delta\chi^2_\text{eff}\approx 20$~\cite{Peiris:2013opa}. Restricting the maximum multipole range for the Planck analysis we do see features in the marginalized likelihood at parameter values corresponding to the best fits in the WMAP analysis of Ref.~\cite{Peiris:2013opa}. However, with $\Delta\chi^2_\text{eff}\approx 10$ the significance of the detection is substantially lower in the Planck data and entirely compatible with cosmic variance, rather than a genuinely physical feature. 

The  Planck and WMAP analyses differ substantially in their details, so there are many possible causes for the difference between the significance of the best fits.  The WMAP analysis uses high-$\ell$ polarization data that is not currently part of the Planck analysis. Repeating the WMAP analysis with temperature data and only low-$\ell$ polarization data  increases the difference in $\Delta \chi^2$, so  this cannot explain the observed difference.  The two analyses also use very different masks. The masks used for the Planck analysis mask a much larger fraction of the sky than those used in the WMAP analysis and the Planck masks are apodized while WMAP masks are not. Finally, Planck and WMAP use different approximations for the likelihood function. The Planck data analysis relies on a Gaussian approximation, while WMAP includes a correction term to take into account the departure from a Gaussian distribution for the $C_\ell$, which are sums of squares of Gaussian random variables and obey a $\chi^2$-distribution. The importance of the non-Gaussian corrections decreases with increasing $\ell$ and the Gaussian approximation for Planck is presumably motivated by the fact that the transition from a pixel-based likelihood to a multipole based likelihood occurs at $\ell=50$ for Planck compared to $\ell=32$ for WMAP. A detailed study investigating the origin of difference in significance between the two experiments will be presented elsewhere.  

With regard to the status of the theoretical predictions, over (most of) the range of axion decay constants we  studied, the single field effective field theory is weakly coupled and its predictions are well understood. Our constraints should thus be thought of as constraints on this single field effective field theory. There are effects in the underlying stringy construction, however, that are not captured by this effective field theory. As inflation continues, the volumes of the internal cycles change due to back-reaction. Since the value of the axion decay constant is set by these volumes, the decay constant will change slightly during inflation.  Consequently, the frequency of the modulation changes slightly as a function of comoving momentum. This effect is less important over the range of scales probed by WMAP, but ideally it should be taken into account for the Planck analysis. Furthermore, the predictions of the stringy model should be worked out more carefully for the shaded regions in our plots in which the predictions of the single field effective field theory start to become unreliable. This is also interesting from a purely theoretical viewpoint as it is not immediately obvious how string theory chooses to UV complete the single field effective field theory as it approaches strong coupling.

Our analysis  focused on monodromy inflation with a linear slow-roll potential. This is merely one among a many monodromy models with potentials that flatten out for large field values relative to a quadratic potential, due to backreaction.  Different potentials are easily implemented, and it would be straightforward to scan over $p$ in $\phi^p$, in addition to the existing set of parameters. However, given the caveats in the previous paragraph, the theoretical predictions  should likely be properly explored before this is done. Consequently, the search for modulations to the primordial power spectrum and in higher n-point functions presents an ongoing challenge for early universe cosmology.\footnote{As this paper was being finished, we became aware of a preprint by Meerburg, Spergel and Wandelt that discusses the problem of searching for oscillatory features in the primordial power spectrum a new approach for doing so, and a second preprint by  Meerburg and Spergel which focuses explicitly on the Planck data. Our results appear to be in broad agreement with these analyses.} Finally we note that the current Planck likelihood may evolve as the full dataset is analyzed, which also has the potential to modify the results presented here.

\section*{Acknowledgments}
We thank Hiranya Peiris for useful discussions. We also thank Daan Meerburg for discussing drafts of his preprints with us. RF is supported by the NSF under grant NSF-PHY-0855425 and NSF-PHY-0645435.  Part of this work was undertaken at the Kavli Institute for Theoretical Physics, which is supported in part by the National Science Foundation under Grant No. NSF PHY11-25915. RF would like to thank the CERN theory division for their hospitality while this work was being completed.


\bibliographystyle{h-physrev3}
\bibliography{modecode4}

\begin{thebibliography}{10}

\bibitem{Ade:2013zuv}
Planck Collaboration, P.~Ade {\em et~al.},
\newblock (2013), 1303.5076.

\bibitem{Ade:2013uln}
Planck Collaboration, P.~Ade {\em et~al.},
\newblock (2013), 1303.5082.

\bibitem{Ade:2013ydc}
Planck Collaboration, P.~Ade {\em et~al.},
\newblock (2013), 1303.5084.

\bibitem{Lyth:1996im}
D.~H. Lyth,
\newblock Phys.Rev.Lett. {\bf 78}, 1861 (1997), hep-ph/9606387.

\bibitem{Linde:1983gd}
A.~D. Linde,
\newblock Phys.Lett. {\bf B129}, 177 (1983).

\bibitem{Linde:1987yb}
A.~D. Linde,
\newblock Phys.Lett. {\bf B202}, 194 (1988).

\bibitem{Freese:1990rb}
K.~Freese, J.~A. Frieman, and A.~V. Olinto,
\newblock Phys.Rev.Lett. {\bf 65}, 3233 (1990).

\bibitem{Silverstein:2008sg}
E.~Silverstein and A.~Westphal,
\newblock Phys.Rev. {\bf D78}, 106003 (2008), 0803.3085.

\bibitem{McAllister:2008hb}
L.~McAllister, E.~Silverstein, and A.~Westphal,
\newblock Phys.Rev. {\bf D82}, 046003 (2010), 0808.0706.

\bibitem{Flauger:2009ab}
R.~Flauger, L.~McAllister, E.~Pajer, A.~Westphal, and G.~Xu,
\newblock JCAP {\bf 1006}, 009 (2010), 0907.2916.

\bibitem{Dong:2010in}
X.~Dong, B.~Horn, E.~Silverstein, and A.~Westphal,
\newblock Phys.Rev. {\bf D84}, 026011 (2011), 1011.4521.

\bibitem{Amin:2011hj}
M.~A. Amin, R.~Easther, H.~Finkel, R.~Flauger, and M.~P. Hertzberg,
\newblock Phys.Rev.Lett. {\bf 108}, 241302 (2012), 1106.3335.

\bibitem{Barnaby:2011qe}
N.~Barnaby, E.~Pajer, and M.~Peloso,
\newblock Phys.Rev. {\bf D85}, 023525 (2012), 1110.3327.

\bibitem{Peiris:2013opa}
H.~Peiris, R.~Easther, and R.~Flauger,
\newblock (2013), 1303.2616.

\bibitem{Bennett:2012fp}
C.~Bennett {\em et~al.},
\newblock (2012), 1212.5225.

\bibitem{Hinshaw:2012fq}
G.~Hinshaw {\em et~al.},
\newblock (2012), 1212.5226.

\bibitem{Mortonson:2010er}
M.~J. Mortonson, H.~V. Peiris, and R.~Easther,
\newblock Phys.Rev. {\bf D83}, 043505 (2011), 1007.4205.

\bibitem{Easther:2011yq}
R.~Easther and H.~V. Peiris,
\newblock Phys.Rev. {\bf D85}, 103533 (2012), 1112.0326.

\bibitem{Feroz:2007kg}
F.~Feroz and M.~Hobson,
\newblock Mon.Not.Roy.Astron.Soc. {\bf 384}, 449 (2008), 0704.3704.

\bibitem{Feroz:2008xx}
F.~Feroz, M.~Hobson, and M.~Bridges,
\newblock Mon.Not.Roy.Astron.Soc. {\bf 398}, 1601 (2009), 0809.3437.

\bibitem{Lewis:2002ah}
A.~Lewis and S.~Bridle,
\newblock Phys.Rev. {\bf D66}, 103511 (2002), astro-ph/0205436.

\bibitem{Liddle:2003as}
A.~R. Liddle and S.~M. Leach,
\newblock Phys.Rev. {\bf D68}, 103503 (2003), astro-ph/0305263.

\bibitem{Peiris:2008be}
H.~V. Peiris and R.~Easther,
\newblock JCAP {\bf 0807}, 024 (2008), 0805.2154.

\bibitem{Adshead:2010mc}
P.~Adshead, R.~Easther, J.~Pritchard, and A.~Loeb,
\newblock JCAP {\bf 1102}, 021 (2011), 1007.3748.

\bibitem{Flauger:2010ja}
R.~Flauger and E.~Pajer,
\newblock JCAP {\bf 1101}, 017 (2011), 1002.0833.

\bibitem{Behbahani:2011it}
S.~R. Behbahani, A.~Dymarsky, M.~Mirbabayi, and L.~Senatore,
\newblock JCAP {\bf 1212}, 036 (2012), 1111.3373.

\bibitem{Cheung:2007st}
C.~Cheung, P.~Creminelli, A.~Fitzpatrick, J.~Kaplan, and L.~Senatore,
\newblock JHEP {\bf 0803}, 014 (2008), 0709.0293.

\bibitem{Meerburg:2011gd}
P.~D. Meerburg, R.~Wijers, and J.~P. van~der Schaar,
\newblock (2011), 1109.5264.

\bibitem{Aich:2011qv}
M.~Aich, D.~K. Hazra, L.~Sriramkumar, and T.~Souradeep,
\newblock (2011), 1106.2798.

\bibitem{Banks:2003sx}
T.~Banks, M.~Dine, P.~J. Fox, and E.~Gorbatov,
\newblock JCAP {\bf 0306}, 001 (2003), hep-th/0303252.

\bibitem{ArkaniHamed:2006dz}
N.~Arkani-Hamed, L.~Motl, A.~Nicolis, and C.~Vafa,
\newblock JHEP {\bf 0706}, 060 (2007), hep-th/0601001.

\bibitem{Planck:2013kta}
Planck collaboration, P.~Ade {\em et~al.},
\newblock (2013), 1303.5075.

\bibitem{cleaning}
R.~Flauger, R.~Hlozek, and D.~Spergel,
\newblock (to appear).

\bibitem{Chen:2008wn}
X.~Chen, R.~Easther, and E.~A. Lim,
\newblock JCAP {\bf 0804}, 010 (2008), 0801.3295.

\bibitem{Behbahani:2012be}
S.~R. Behbahani and D.~Green,
\newblock JCAP {\bf 1211}, 056 (2012), 1207.2779.

\end{thebibliography}


\end{document}